\documentclass[11pt,onecolumn,amssymb,nofootinbib]{revtex4}
\usepackage{amsmath, amsthm, amscd, amssymb}
\usepackage{bm}
\usepackage{bbm}

\begin{document}

\title{\bf Minimum Interior Temperature for Solid Objects Implied by Collapse Models}

\author{Stephen L. Adler}
\email{adler@ias.edu} \affiliation{Institute for Advanced Study,
Einstein Drive, Princeton, NJ 08540, USA.}

\begin{abstract}
Heating induced by the noise postulated in wave function collapse models leads to a lower bound to the temperature of solid objects.  For the noise parameter values
$\lambda ={\rm coupling~strength}\sim 10^{-8} {\rm s}^{-1}$ and $r_C ={\rm correlation~length} \sim 10^{-5} {\rm cm}$, which were suggested \cite{adler1} to make latent image formation an indicator of wave
function collapse and which are consistent with the recent
experiment of Vinante et al. \cite{vin}, the effect may be observable.   For metals, where the heat conductivity is proportional to the temperature at low temperatures, the lower bound (specifically for RRR=30 copper) is  $\sim 5\times 10^{-11} (L/r_C) $K, with L the size of the object. For the thermal insulator
Torlon 4203, the comparable lower bound is $\sim 3 \times 10^{-6} (L/r_c)^{0.63}$ K.
 We first give a rough estimate for a cubical metal solid, and then give  an exact solution of the heat transfer problem for a sphere.

\end{abstract}
\maketitle

There is increasing interest in testing wave function collapse models \cite{collapse},  by searching for effects associated with the small noise which drives wave function collapse when nonlinearly coupled in the Schr\"odinger
equation.  The original proposals for the noise coupling strength were so small that devising
suitable experiments was problematic, but the situation has changed with the suggestion \cite{adler1}   that latent image formation, such as deposition of a developable track in an emulsion or in an etched track detector, already constitutes a measurement embodying wave function collapse. A recent
cantilever experiment of Vinante et al. \cite{vin}  has set bounds consistent with the enhanced parameters suggested in \cite{adler1}, and reports a possible noise signal. Thus, it is timely
to consider other experiments \cite{alternative} which could detect or rule out a noise coupling with
the strength suggested by \cite{vin}.

For a body comprised of a group of particles of total mass $M$, the secular center-of-mass energy gain is given by the formula \cite{gain}
\begin{equation}\label{engain}
\frac{dE}{dt}=\frac{3}{4}\lambda \frac{\hbar^2}{r_C^2} \frac{M}{m_N^2}~~~,
\end{equation}
with $m_N$ the nucleon mass.  For a body of dimensions $L$ larger than the correlation length
$r_C$, the different groups of particles will have independent center of mass motions, and
so the energy gain from Eq. \eqref{engain} will take the form of thermal energy.  With the
stipulation that we will only be considering bodies of uniform mass density $\rho$ with dimensions $L>>r_C$, we rewrite Eq. \eqref{engain} as a formula for the energy deposition
rate per $Q$  unit volume given as
\begin{equation}\label{qform}
Q=\frac{3}{4}\lambda \frac{\hbar^2}{r_C^2} \frac{\rho}{m_N^2}~~~.
\end{equation}

For an initial estimate, consider a solid metal cube of side length $L$, with heat conductivity  $k(T)$, which at  temperatures $T$ below $\sim 10\,{\rm K}$ obeys the linear law  $k(T) = k_0 T$.  The specific example
that we shall use for an estimate is medium purity Residual Resistivity Ratio (RRR) $=30$  copper, for which \cite{kest} $k_0 \simeq 45 {\rm W}/{\rm (m ~K^2)}=45 {\rm Joule}/{\rm (m ~K^2~s)}$.
Assuming the body has surface temperature $T_s=0$ and central temperature $T_c$, the rate at
which energy is transported out by conduction is approximately
\begin{equation}\label{enout}
E_{\rm out}=6 L^2 (k_0 T_c/2) (T_c/L) =3 L k_0 T_c^2 ~~~.
\end{equation}
At equilibrium, this must balance the rate of noise-induced heating given by
\begin{equation}\label{enin}
E_{\rm in}=Q L^3~~~.
\end{equation}
Equilibrium can be attained in a reasonably short time since the thermal diffusivity of metals, which is proportional to the ratio of the heat conductivity $k$ to the specific heat capacity
$c_p$ times the density, increases as the temperature decreases \cite{kest}, because $c_p$ decreases more rapidly with decreasing temperature than does $k$.  Equating $E_{\rm out}$ with $E_{\rm in}$ gives
a formula for the central temperature $T_c$,
\begin{equation}\label{tcform}
T_c=\theta \frac{L}{r_c} {\rm K}~~~,
\end{equation}
with $\theta$ the dimensionless number
\begin{equation}\label{thetdef}
\theta=\left[\frac{\lambda \hbar^2 \rho}{4 (k_0{\rm K}^2) m_N^2}\right]^{1/2}~~~.
\end{equation}
Taking the values (for RRR=$30$ copper)
\begin{align}\label{values}
\lambda =& 10^{-8} {\rm s}^{-1}~~~,\cr
\hbar=& 1.1 \times 10^{-34} {\rm Joule} ~{\rm s}~~~,\cr
\rho=& 9 ~{\rm gm} ~{\rm cm}^{-3} = 5.0 \times 10^{27} ~({\rm MeV}/c^2) ~{\rm cm}^{-3}~~~,\cr
k_0{\rm K}^2=&45~ {\rm Joule}/({\rm m~s})~~~,\cr
m_N=& 940 ~{\rm MeV}/c^2 ~~~,\cr
c=&3 \times 10^{10} {\rm cm}/{\rm s}~~~,\cr
1~{\rm Joule}=& 6.2 \times 10^{12} {\rm MeV}~~~,\cr
\end{align}
we find
\begin{equation}\label{thetvalue}
\theta\simeq 4.6 \times 10^{-11}~~~.
\end{equation}
For $L=1 {\rm m}$, Eq. \eqref{tcform} gives $T_c  \sim 5 \times 10^{-4}$ K.  For comparison, 400 kg of copper has been cooled \cite{copper} to $6 \times 10^{-3}$ K, not constraining
the $\lambda$ value used in this $T_c$ estimate.

A  spin temperature of $0.1 \times 10^{-9}$ K and a lattice temperature of $6 \times 10^{-5}$ K have been reported \cite{rhodium} for a $0.4 \times 4 \times 25~ {\rm mm}^3$ rhodium single crystal.  Taking the smallest dimension $0.4$ mm as the relevant $L$ for an estimate, Eq. \eqref{tcform} gives $T_c \simeq 4 \times 10^{-7}$ K.   Thus assuming that this estimate should be compared to the lattice temperature and not the spin temperature (an assumption that deserves further study), the rhodium experiment also does not give a constraint  on the noise coupling parameter $\lambda$. Cooling of a drum-shaped aluminum membrane 20 microns wide and 100 nm thick has been reported \cite{alum} to the
temperature $4 \times 10^{-4}$ K.   Again taking the minimum dimension, which in this case is $\sim r_c$, as the relevant L, this experiment also does not give a constraint on $\lambda$.

To give a more precise estimate of $T_c$ and of the temperature variation throughout the
volume of a solid body, we consider the simplest case of a sphere of radius $L$ with thermal conductivity $k(T)$ and heating rate $Q$.  At a given distance $R$ from the center of the
sphere, the energy transport rate thorough the spherical surface of radius R is equal to
\begin{equation}\label{spherout}
E_{\rm out}= -4 \pi R^2 k(T) \frac{dT}{dR}~~~,
\end{equation}
which at equilibrium must balance the heating rate of the volume within radius $R$,
\begin{equation}\label{spherin}
E_{\rm in}=\frac{4 \pi}{3} R^3 Q~~~,
\end{equation}
giving the differential equation
\begin{equation}\label{diffeq}
-k(T)\frac{dT}{dR}=\frac{1}{3} R Q~~~.
\end{equation}
Integrating from the center of the sphere at radius 0  to radius $R$, this gives
\begin{equation}\label{int}
-\int_{T_c}^T \, k(u)du= \frac{R^2 Q}{6}~~~.
\end{equation}
For $k(u)=\hat k_0 u^{\beta}$,
this becomes
\begin{equation}\label{int1}
-\frac{\hat k_0}{1+\beta} (T^{1+\beta}-T_c^{1+\beta})=\frac{R^2 Q}{6}~~~.
\end{equation}

If we assume a boundary condition $T=T_s$ at the outer surface of the sphere at $R=L$, Eq. \eqref{int1} implies
that
\begin{equation}\label{intt1}
 (T_c^{1+\beta}-T_s^{1+\beta})=\frac{1+\beta}{\hat k_0} \frac{L^2 Q}{6}~~~,
\end{equation}
which gives the inequality
\begin{equation}\label{ineq1}
T_c = \left[T_s^{1+\beta}+\frac{1+\beta}{\hat k_0}\frac{L^2 Q}{6} \right]^{1/(1+\beta)}\geq \left[\frac{1+\beta}{\hat k_0}\frac{L^2 Q}{6} \right]^{1/(1+\beta)}~~~,
\end{equation}
so that on inserting Eq. \eqref{qform} we get
\begin{equation}\label{ineq2}
T_c \geq \left[\frac{1+\beta}{(\hat k_0{\rm K}^{1+\beta})}\frac{\lambda \hbar^2 \rho}{8 m_N^2}\right]^{1/(1+\beta)} \left(\frac{L}{r_c}\right)^{2/(1+\beta)} {\rm K}~~~.
\end{equation}
Specializing to a metal with $\beta=1$ and $\hat k_0=k_0$, Eq. \eqref{ineq2} reproduces Eqs. \eqref{tcform} and \eqref{thetdef}.  For the
thermal insulator Torlon 4203 \cite{torlon}, with density $1.42~ {\rm gm~cm^{-3}}$ and with $k(T)=6.13 \times 10^{-3} (T/{\rm K})^{2.18} {\rm W/(m~K)}$, so that $\beta=2.18$ and $\hat k_0 {\rm K}^{1+\beta}= 6.13 \times 10^{-3}$, Eq. \eqref{ineq1} gives $T_c \geq 3.4 \times 10^{-6} (L/r_c)^{0.63}$ K.  For example, for a Torlon sphere
of radius $L=50\,{\rm cm}$, the central temperature $T_c=5.6 \times 10^{-2} \,{\rm K}$, nearly a factor of 10 bigger than the temperature attained in \cite{copper} using the
CUORE experiment cryostat.\footnote{Taking $\lambda=10^{-7.7} {\rm s}^{-1}$ as suggested in \cite{vin}, instead of the nominal value $\lambda=10^{-8} {\rm s}^{-1}$ used in our estimates, increases
this by a factor $10^{0.3/3.18}=1.24$  to $T_c=7.0\times 10^{-2} \,{\rm K}$.}  This shows that with thermal
insulating material arranged in a compact geometry, such as a sphere or cube, the effect we are proposing could
be detected using the CUORE cryostat.  In the actual running of the CUORE experiment, the bolometers consisted
of $5 \times 5 \times 5 ~{\rm cm}^3$ cubes of ${\rm TeO}_2$ stacked in 19 towers, so all (except those at the
ends) are in the same geometry relative to the cyostat.   With this configuration of material, one would not
expect to see a dramatic difference between internal and surface temperatures.

Equations \eqref{int}--\eqref{intt1} can be used to give the temperature profile in the sphere as a function of radius $R$. For a cylinder of infinite length, the same equations apply with the
substitution $Q \to 3Q/2$, with radii now referring to the cylinder.  To get the temperature profile for other
geometries of interest, such as an ellipsoid of revolution, a cylinder of finite length, or a rectangular parallelepiped, one must solve the nonlinear differential equation governing thermal equilibrium
\begin{equation}\label{equildiff}
- \vec \nabla \cdot \big( k(T) \vec \nabla T\big)=Q~~~
\end{equation}
with suitable boundary conditions.  When $k(T)=\hat k_0 T^{\beta}$, using $ T^{\beta} \vec \nabla T = (1+\beta)^{-1}\vec \nabla T^{1+\beta}$, Eq.
\eqref{equildiff} becomes the Poisson equation
\begin{equation}\label{equildiff1}
{\vec \nabla}^2 T^{1+\beta}+ \frac{(1+\beta) Q}{\hat k_0}=0~~~,
\end{equation}
which can be solved by standard  methods \cite{morse}.   

To conclude, the decrease in thermal conductivity at low temperatures for both metals and thermal insulators results in a ``trapped heat''
phenomenon, in which the noise-induced heating associated with collapse models results in lower bounds on the internal temperature of solid objects.
The fact that these lower bounds scale up with increasing $L/r_c$ may make experiments to search for them feasible.  Clearly  measuring the central temperature $T_c$ of a solid object without disturbing its thermal equilibrium will be a technical challenge.  For larger objects, sensors with fine wire leads could be used.  For smaller
objects, central temperatures could be probed using a small hole from the exterior
to the cental region, through which molecules of an evaporative medium placed at the center can pass, or through which a laser beam can
be directed to detect the state of molecular motion at the center.
In designing experiments,
it will be important to make sure that the volumetric heating per unit mass from radioactivity and particle penetration  is significantly less than
\begin{equation}\label{engain1}
\frac{dE}{dt\,dM}=\frac{3}{4}\lambda \frac{\hbar^2}{r_C^2} \frac{1}{m_N^2}\simeq 20 \frac{{\rm MeV}}{{\rm gm ~s}}~~~,
\end{equation}
where we have again used the parameter values of Eq. \eqref{values}.  This brings up an important caution underlying our analysis, which is that we have assumed that the energy production rate of Eqs. \eqref{engain} and \eqref{engain1} applies to laboratory scale objects.  As already noted in \cite{adler1}, this assumption breaks down when applied to the Earth's heat flow, where assuming an energy production rate of Eq. \eqref{engain1} througout the interior of the Earth leads to an internal heat production
roughly three orders of magnitude larger than what is observed.  However, as also noted in \cite{adler1}, when the effects of dissipation are included, as in the model of Bassi, Ippoliti, and Vachini \cite{ippoliti}, the rate of heat production can vanish at large times where a limiting temperature is reached. For example, with the parameters of \cite{ippoliti},  a limiting temperature of $0.1$ K is reached on a time scale of billions of years, indicating a
current noise-induced Earth  heat production rate much smaller than given
by naive application of Eq. \eqref{engain1}.  Further study of this issue is warranted.

I wish to thank Angelo Bassi for helpful comments.


\begin{thebibliography}{99}
\bibitem{adler1} S. L. Adler, J. Phys. A: Math. Theor. {\bf 40}, 2935, (E) 13501 (2007).
\bibitem{vin} A. Vinante, R. Mezzena, P. Falferi, M Carlesso, and A. Bassi,  Phys. Rev. Lett. {\bf 119}, 110401 (2017).
\bibitem{alternative}  Alternative recent proposals include:  M. Bahrami, M. Paternostro, A. Bassi, and H. Ulbricht,  Phys. Rev. Lett. {\bf 112}, 210404 (2014);
S. Nimmrichter, K. Hornberger, and K. Hammerer, Phys. Rev. Lett. {\bf 113}, 020405 (2014); F. Lalo\"e, W. J. Mullin, and P. Pearle, Phys. Rev. A {\bf 90}, 052119 (2014); 
L. Di\'osi, Phys. Rev. Lett. {\bf 114}, 050403 (2015); D. Goldwater,
M. Paternostro, and P. F. Barker, Phys. Rev. A {\bf 94}, 010104 (2016).
\bibitem{collapse} For reviews see:  A. Bassi and G. C. Ghirardi, Phys. Rep. {\bf 379}, 257 (2003);  P. Pearle, in ``Open Systems and Measurements in Relativistic
Quantum Field Theory'', Lecture Notes in Physics Vol.  526, H.-P. Breuer and F. Petruccione, eds., Springer, Berlin, 1999.
\bibitem{gain}  P. Pearle and E. Squires, Phys. Rev. Lett. {\bf 73}, 1 (1994), Eq. (2); S. L. Adler, ref [1] op. cit., Eq. (7) and the related comments contained in references [5] and [6]. For a detailed derivation in the continuous spontaneous localization (CSL) model, see F. Lalo\"e, W. J. Mullin, and P. Pearle, ref [3] op. cit., Appendix A.
\bibitem{kest} P. Duthil, ``Material Properties at Low Temperatures'', arXiv:1501.07100.
\bibitem{copper} CUORE collaboration press release (2014); see https://www.interactions.org/press-release/infn-cuore.
\bibitem{rhodium} T. Knuuttila, ``Nuclear Magnetism and Superconductivity in Rhodium'',  Helsinki University
of Technology thesis (2000), https://web.archive.org/web/20110816203628/http://lib.tkk.fi/Diss/2000/isbn9512252147/isbn9512252147.pdf.  For the
physical properties of rhodium, see: http://periodictable.com/Elements/045/data.html.
\bibitem{alum} See: http://www.iflscience.com/physics/coldest-temperature-universe-created-american-laboratory/ (2017).
\bibitem{torlon} G. Ventura et al., Cryogenics {39}, 481 (1999).  For the density and other properties of torlon:
https://www.professionalplastics.com/professionalplastics/content/downloads/Solvay\_Torlon\_Design\_Guide.pdf
\bibitem{morse} P. M. Morse and H. Feshbach, ``Methods of Theoretical Physics'', McGraw-Hill, New York (1953), Chapter 10.  
\bibitem{ippoliti} A. Bassi, E. Ippoliti, and B. Vacchini, J. Phys. A: Math Gen. {\bf 38}, 8017 (2005).  See A. Smirne and A. Bassi, Sci. Rep. {\bf 5}, 12518 (2015) for 
the extension to the dissipative CSL model.  
\end{thebibliography}
\end{document}